\documentclass[%
reprint,
superscriptaddress,
 amsmath,amssymb,
 aps,
floatfix,
]{revtex4-2}

\usepackage{graphicx}% Include figure files
\usepackage{dcolumn}% Align table columns on decimal point
\usepackage{bm}% bold maths
\usepackage{hyperref}% add hypertext capabilities

\usepackage{aas_macros}

\usepackage{psfrag}
\usepackage{pstool}
\usepackage{pstricks}
\usepackage{bm}
\usepackage{pgfplots}
\pgfplotsset{compat=newest}
\usepackage{tikz-3dplot}
    \usetikzlibrary{decorations.markings,cd,calc,patterns,angles,quotes,shapes,plotmarks, decorations.pathmorphing,positioning, arrows,backgrounds}

    \tikzset{>=stealth}
    \pgfplotsset{every axis/.style={scale only axis}}
\usepackage{wrapfig}
\usepackage{mathtools}
\usepackage{epstopdf}

\definecolor{MATLABMagenta}{RGB}{255, 0, 255}

\usepackage{stackengine}
\stackMath

\DeclarePairedDelimiter\abs{\lvert}{\rvert}%
\DeclarePairedDelimiter\norm{\lVert}{\rVert}%

\makeatletter
\let\oldabs\abs
\def\abs{\@ifstar{\oldabs}{\oldabs*}}
\let\oldnorm\norm
\def\norm{\@ifstar{\oldnorm}{\oldnorm*}}
\makeatother

\begin{document}

\preprint{APS/123-QED}

\title{Self-stabilization of light sails by damped internal degrees of freedom}

\author{M. Z. Rafat}
\email{mohammad.rafat@sydney.edu.au}
\affiliation{School of Physics, University of Sydney, Sydney, NSW 2006, Australia}

\author{Holger R. Dullin}
\affiliation{School of Mathematics and Statistics, University of Sydney, Sydney, NSW 2006, Australia}

\author{Boris T. Kuhlmey}
\affiliation{School of Physics, University of Sydney, Sydney, NSW 2006, Australia}

\author{Alessandro Tuniz}
\affiliation{School of Physics, University of Sydney, Sydney, NSW 2006, Australia}

\author{Haoyuan Luo}
\affiliation{School of Physics, University of Sydney, Sydney, NSW 2006, Australia}

\author{Dibyendu Roy}
\affiliation{School of Physics, University of Sydney, Sydney, NSW 2006, Australia}

\author{Sean Skinner}
\affiliation{School of Physics, University of Sydney, Sydney, NSW 2006, Australia}

\author{Tristram J. Alexander}
\affiliation{School of Physics, University of Sydney, Sydney, NSW 2006, Australia}

\author{Michael S. Wheatland}
\affiliation{School of Physics, University of Sydney, Sydney, NSW 2006, Australia}

\author{C. Martijn de Sterke}
\affiliation{School of Physics, University of Sydney, Sydney, NSW 2006, Australia}

% \collaboration{MUSO Collaboration}%\noaffiliation

% \affiliation{
% 	Second institution and/or address\\
% 	This line break forced% with \\
% }%
% \affiliation{
% 	Third institution, the second for Charlie Author
% \affiliation{%
% 	Authors' institution and/or address\\
% 	This line break forced with \textbackslash\textbackslash
% }%

% \collaboration{CLEO Collaboration}%\noaffiliation

\date{\today}% It is always \today, today,

\begin{abstract}

%Please use \texttt{\textbackslash myEdit\{edited text\}} and \texttt{\textbackslash myComment\{your comment\}\{text in question\}}.\\\\

We consider the motion of a light sail that is accelerated by a powerful laser beam. We derive the equations of motion for two proof-of-concept sail designs with damped internal degrees of freedom. Using linear stability analysis we show that perturbations of the sail movement in all lateral degrees of freedom can be damped passively. This analysis also shows complicated behaviour akin to that associated with exceptional points in PT-symmetric systems in optics and quantum mechanics. The excess heat that is produced by the damping mechanism is likely to be substantially smaller than the expected heating due to the partial absorption of the incident laser beam by the sail.

\end{abstract}

\maketitle

%\tableofcontents

\section{\label{sec:introduction} Introduction}

The task of sending spacecrafts to planets and other objects within our own solar system has become almost routine. The Breakthrough Starshot Initiative aims to expand our horizons beyond our own solar system to our closest neighbour, the Alpha Centauri system at a distance of 4.2 light years from Earth \cite{BSI_website}. This enormous distance means that even the fastest man-made craft to date, the Parker Solar Probe (with a predicted top speed of 0.064\% of the speed of light at its closest approach to the Sun \cite{PSP_document}), would take about 6500 years to reach Alpha Centauri. A spacecraft accelerated by chemical fuel needs to carry a prohibitively large amount of fuel in order to reach speeds nearing any appreciable fraction of the speed of light. A natural candidate for an external source of energy is light, which was proposed decades ago \cite{Marx_1966_1966Natur.211...22M,Forward_1984_1984JSpRo..21..187F}. This is the rationale for the approach taken by the Breakthrough Starshot Initiative. The aim is to send an ultralight sail craft with a payload to Alpha Centauri by accelerating it to about 20\% of the speed of light using an Earth-based laser phased array \cite{Lubin_2016_2016JBIS...69...40L}. This would allow the sail to reach Proxima Centauri and send signals back to Earth in about 26 years; all within a human lifetime. The sail is expected to have a mass of about one gram, with a payload that contains the detectors and electronics necessary to send a signal back to Earth having a similar mass \cite{Atwater_etal_2018_2018NatMa..17..861A}. There are numerous scientific and engineering challenges in every aspect of this grand vision, including laser array design \cite{Bandutunga_etal_2021_Bandutunga:21}, material choice \cite{Atwater_etal_2018_2018NatMa..17..861A, Davoyan_etal_2021_2021Optic...8..722D}, stability of the sail under acceleration \cite{Ilic_Atwater_2019_2019NaPho..13..289I}, heat management \cite{Atwater_etal_2018_2018NatMa..17..861A, Jin_etal_2020_Jin_2020, Ilic_etal_2018_2018NanoL..18.5583I} and communication \cite{Messerschmitt_etal_2020_2020ApJS..249...36M}.

Diffraction of the accelerating beam and size limits on the laser array imply that a ``reasonable'' scheme to accelerate the sail to its final speed is as follows \cite{Lubin_2016_2016JBIS...69...40L}: The total area of the sail is $ \sim 10~{\rm m^2}$ and the net incoming laser intensity is approximately $ 10~{\rm GW\,m^{-2}} $. The sail is accelerated to approximately $ 20\% $ of the speed of light over a distance of approximately $ 10 $ times the distance between Earth and the Moon in about $ 15 $ minutes. To increase momentum transfer, the sail material needs to have high refractive index, potentially in multiple layers \cite{Ilic_etal_2018_2018NanoL..18.5583I, Jin_etal_2020_Jin_2020}, with low absorption in the laser wavelength range and high absorption/emissivity at longer wavelengths to cool the sail through thermal radiation. The large surface area and low mass implies that the material must have low mass density \cite{Atwater_etal_2018_2018NatMa..17..861A}. 

\begin{figure*}
    \centering
    \begin{tikzpicture}
        \node[anchor=south west,inner sep=0] (image) at (0,0) {\includegraphics[width=\textwidth]{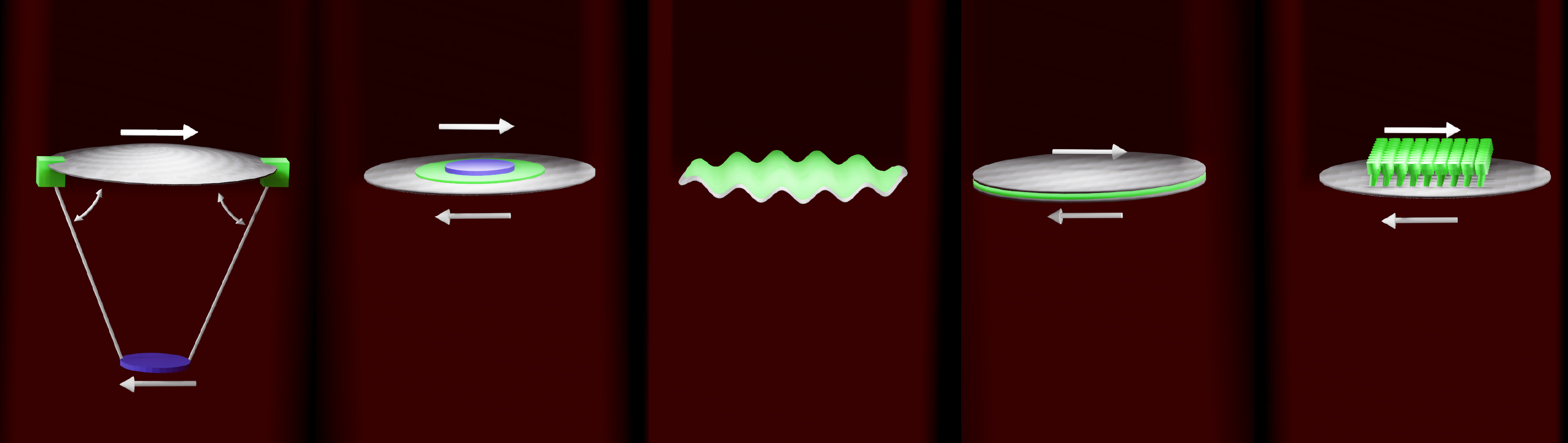}};
        \definecolor{myBG}{rgb}{1,1,1}
        \begin{scope}[x={(image.south east)},y={(image.north west)}]
            \node[rectangle,draw=white,fill=myBG,minimum size=1.5em,label={center:(a)}] at (0.03, 0.9) {};
            \node[rectangle,draw=white,fill=myBG,minimum size=1.5em,label={center:(b)}] at (0.25, 0.9) {};
            \node[rectangle,draw=white,fill=myBG,minimum size=1.5em,label={center:(c)}] at (0.45, 0.9) {};
            \node[rectangle,draw=white,fill=myBG,minimum size=1.5em,label={center:(d)}] at (0.65, 0.9) {};
            \node[rectangle,draw=white,fill=myBG,minimum size=1.5em,label={center:(e)}] at (0.85, 0.9) {};
        \end{scope}
    \end{tikzpicture}
    \caption{Examples of conceivable implementations of damped internal degrees of freedom for lightsails. Features are not to scale. Green: elastic damped medium; blue: payload (when not distributed on sail); Red: laser beam. (a) damped hinges; (b) central payload attached by viscoelastic medium; (c) damped modes of vibration; (d) dual layers separated by viscoelastic medium; (d) distributed damped cantilever. Arrows: examples of damped transverse motion.}
    \label{fig:damping_schemes}
\end{figure*}
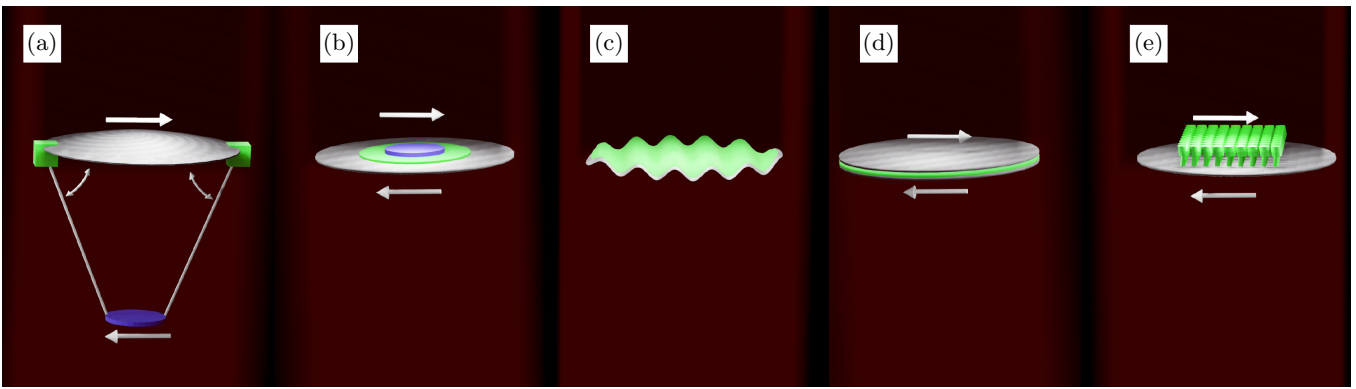

The severe limitation on the total mass of the craft precludes active stabilisation and the enormous acceleration, about 66,000 times that due to Earth's gravity, experienced by the sail implies short dynamical timescales that do not allow feedback loops for remote correction of the sail trajectory through adjustment of the laser beam profile. Instead, the craft needs to passively correct any perturbations that may cause it to exit the accelerating laser beam. A number of rigid body sail designs have been proposed that passively produce restoring forces on the sail towards the centre of the beam \cite{Atwater_etal_2018_2018NatMa..17..861A, Srivastava_etal_2019_2019OptL...44.3082S,Salary_Mosallaei_2020_2020LPRv...1400311S,Manchester_Loeb_2017_2017ApJ...837L..20M,Siegel_etal_2019_Siegel:19, Popova_etal_2017_https://doi.org/10.1002/mma.4282, Schamiloglu_etal_2001_2001AIPC..552..559S, Myilswamy_etal_2020_2020OExpr..28.8223M}. These designs employ variations in geometry (e.g. flat, conical, parabolic and spherical), materials (various semi-conducting materials) and surface structure (specular reflection and various types of metasurfaces) to manipulate the direction of the force of the laser light on the sail. A salient feature in all of these designs is that these restoring forces are undamped and lead to oscillations of the orientation and displacement of the sail perpendicular to the laser beam. We refer to these collectively as \textit{lateral oscillations} of the sail. These undamped lateral oscillations are problematic as they can potentially grow due to nonlinear coupling and/or if additional perturbations are imparted on the sail, for example due to inevitable deviations of the accelerating laser beam from an ideal shape \cite{Bandutunga_etal_2021_Bandutunga:21}. To overcome this problem a damping term was added by Srivastava \textit{et al.} \cite{Srivastava_etal_2019_2019OptL...44.3082S} to the equations of motion, whereas Salary and Mosallaei \cite{Salary_Mosallaei_2020_2020LPRv...1400311S} discuss potential damping of lateral motion of the sail through Doppler damping. However, Srivastava \textit{et al.} do not discuss a physical implementation, and the mechanism and effectiveness of Doppler damping require further study.

We propose and investigate an alternative approach which consists of incorporating damped internal degrees of freedom in the sail. The aim is to couple the lateral oscillations of the sail to the damped internal degrees of freedom, causing the lateral oscillations to decay. This principal has been used extensively in many areas of engineering including damping of oscillations of high rise buildings caused by earthquakes and high winds \cite{Kitagawa_etal_1998_1998SMaS....7..581K, Aly_2014_https://doi.org/10.1002/tal.1068}, vibration of racing cars (Renault R26) and spacecrafts (NASA's Ares solid fuel booster). In a lightsail, incorporating a damped internal degree of freedom could be challenging, and the implementation would depend on the exact sail geometry, which is still being debated. In a geometry such as IKAROS \cite{IKAROS_website}, which has several sails connected by a frame, one could have damped hinges on the frame, Fig.~\ref{fig:damping_schemes}(a). For the Starshot project, most lightsails proposed have been modelled as flat rigid sheets with a central payload chip or a distributed payload. A central payload could be attached to the chip through an elastic and damped connection, Fig.~\ref{fig:damping_schemes}(b), with the relative motion between sail and payload being the damped internal degree(s) of freedom. For distributed payloads the elasticity of the sail itself could be used as damped internal degrees of freedom, Fig.~\ref{fig:damping_schemes}(c), with damping coming for example from passive magnetic induction or phonon scattering. One could even imagine multiple sheets connected by elastoviscous media with higher damping, Fig.~\ref{fig:damping_schemes}(d), or distributed damped cantilevers, Fig.~\ref{fig:damping_schemes}(e). 

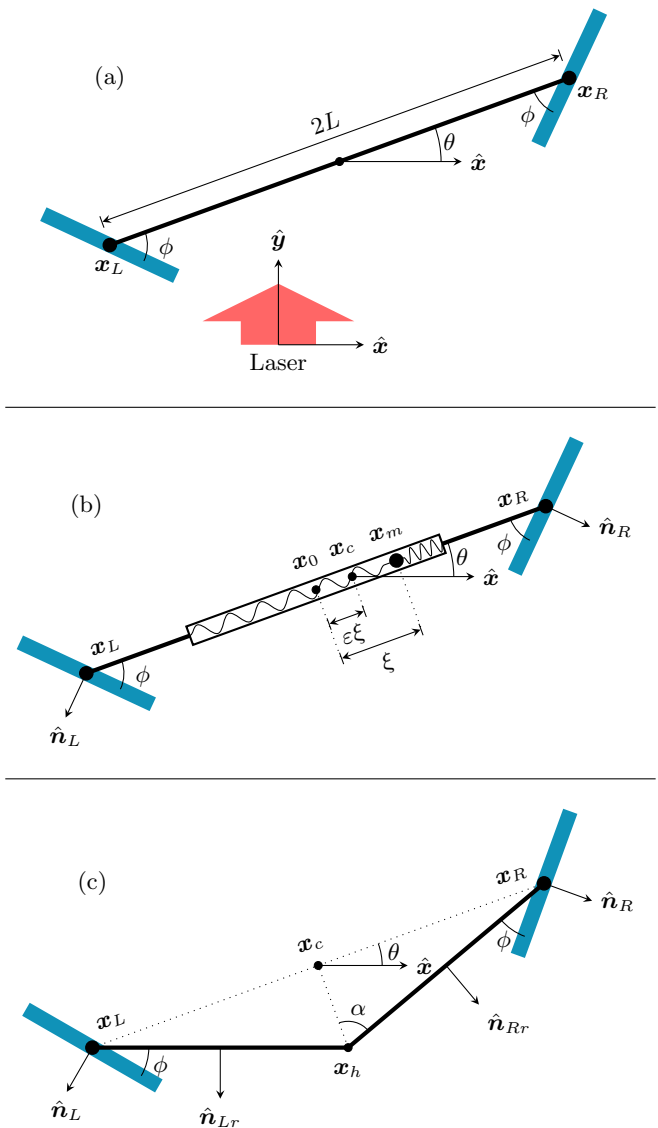
\begin{figure}[hbt]

\centering
    
\pgfmathsetmacro{\lL}{5}
\pgfmathsetmacro{\lLt}{2.75}
\pgfmathsetmacro{\lxi}{1.75}
\pgfmathsetmacro{\lVarepsilon}{0.45}
\pgfmathsetmacro{\aAlpha}{70}
\pgfmathsetmacro{\aTheta}{20}
\pgfmathsetmacro{\aPhi}{45}
\pgfmathsetmacro{\bPhi}{30}

\begin{tikzpicture}[scale=0.65]
    % left mirror coordinates
    \coordinate (xL) at ({-\lL*cos(\aTheta)}, {-\lL*sin(\aTheta)});
    
    % right mirror coordinates
    \coordinate (xR) at ({\lL*cos(\aTheta)}, {\lL*sin(\aTheta)});
    
    % origin node
    \node[scale=0.75, black] (orig) at (0,0) {\pgfuseplotmark{*}};

    % left mirror
    \draw[draw=cyan!70!black, line width=2mm] ($(xL)-1.5*({-cos(\aTheta - \aPhi)}, {-sin(\aTheta - \aPhi)})$) -- ($(xL)+1.5*({-cos(\aTheta - \aPhi)}, {-sin(\aTheta - \aPhi)})$);
    
    % right mirror
    \draw[draw=cyan!70!black, line width=2mm] ($(xR)-1.5*({cos(\aTheta + \aPhi)}, {sin(\aTheta + \aPhi)})$) -- ($(xR)+1.5*({cos(\aTheta + \aPhi)}, {sin(\aTheta + \aPhi)})$);

    % rod
    \draw[-,black, ultra thick] (xL) -- (xR);

    % C coordinates
    \draw[->] (0,0) -- ++({0.5*\lL},0);
        \node[anchor=west] (Cx) at ($(0,0) + 0.5*(\lL, 0)$) {$ \hat{\bm{x}} $};

    % left mirror node
    \node[scale=1.25] (Lm) at (xL) {\pgfuseplotmark{*}};
        \node[anchor=north, inner sep=2.0mm, rotate=0] at (xL) {$ \bm{x}_L $};
        
    % right mirror node
    \node[scale=1.25] (Rm) at (xR) {\pgfuseplotmark{*}};
        \node[anchor=north west, inner sep=1.0mm, rotate=0] at (xR) {$ \bm{x}_R $};
    
    % phi left mirror node
    \node (Lphi) at ($(xL)-1.0*({-cos(\aTheta - \aPhi)}, {-sin(\aTheta - \aPhi)})$) {};
    
    % phi right mirror node
    \node (Rphi) at ($(xR)-1.0*({cos(\aTheta + \aPhi)}, {sin(\aTheta + \aPhi)})$) {};
    
    % Angle markers
    \pic [draw, -, "$\theta$", angle eccentricity=1.1, angle radius=1.35cm] {angle = Cx--orig--Rm};
    \pic [draw, -, "$\phi$", angle eccentricity=1.5, angle radius=0.5cm] {angle = orig--xR--Rphi};
    \pic [draw, -, "$\phi$", angle eccentricity=1.5, angle radius=0.5cm] {angle = Lphi--xL--orig};
    
    % Laser
    \draw[-{Triangle[width=20mm,length=5mm]}, line width=10mm,red!60!white] ($(orig) + (-0.25*\lL, {-0.75*\lL})$) -- ++(0, 1.25);
    \node[anchor=north, rectangle] (laserNode) at ($(orig) + (-0.25*\lL, {-0.75*\lL})$) {Laser};
    
    \draw[->] ($(orig) + (-0.25*\lL, {-0.75*\lL})$) -- ++({0.35*\lL},0) node[anchor=west] {$ \hat{\bm{x}} $};
    \draw[->] ($(orig) + (-0.25*\lL, {-0.75*\lL})$) -- ++(0, {0.35*\lL}) node[anchor=south] {$ \hat{\bm{y}} $};
    
    \draw [|<->|] ($(xL) - 0.5*({sin(\aTheta)},{-cos(\aTheta)})$) -- ($(xR) - 0.5*({sin(\aTheta)},{-cos(\aTheta)})$)  node[midway, anchor = south,rotate=\aTheta] {$ 2 L $};

    \definecolor{myBG}{rgb}{1,1,1}
    \node[rectangle,draw=white,fill=myBG,minimum size=2em,label={center:(a)}] at ({-\lL*cos(\aTheta)}, {\lL*sin(\aTheta)}) {};
\end{tikzpicture}
\vspace{1em}
\hrule
\vspace{1em}
%
% \begin{subfigure}[b]{\columnwidth}
\begin{tikzpicture}[scale=0.65]

    % left mirror
    \coordinate (xL) at ({-\lL*cos(\aTheta)}, {-\lL*sin(\aTheta)});
    
    % right mirror
    \coordinate (xR) at ({\lL*cos(\aTheta)}, {\lL*sin(\aTheta)});
    
    % moveable mass
    \coordinate (xM) at ({\lxi*cos(\aTheta)}, {\lxi*sin(\aTheta)});
    
    % centre of mass
    \coordinate (xC) at ($\lVarepsilon*(xM)$);
    
    % left container edge
    \coordinate (xLC) at ({-\lLt*cos(\aTheta)}, {-\lLt*sin(\aTheta)});
    
    % right container edge
    \coordinate (xRC) at ({\lLt*cos(\aTheta)}, {\lLt*sin(\aTheta)});
    
    % container
    \draw [-, black, thick, rotate around={\aTheta:(xLC)}] ($(xLC)+(0,-0.20)$) rectangle ++({2*\lLt}, 0.40);

    % left spring
    \draw[black, decoration={aspect=1, segment length=4.5mm, amplitude=1mm,snake},decorate] (xLC) -- ($0.99*(xM)$);
    
    % right spring
    \draw[black, decoration={aspect=1, segment length=1.5mm, amplitude=1mm,snake},decorate] (xRC) -- ($1.055*(xM)$);
    
    % left mirror
    \draw[draw=cyan!70!black, line width=2mm] ($(xL)-1.5*({-cos(\aTheta - \aPhi)}, {-sin(\aTheta - \aPhi)})$) -- ($(xL)+1.5*({-cos(\aTheta - \aPhi)}, {-sin(\aTheta - \aPhi)})$);
    
    % right mirror
    \draw[draw=cyan!70!black, line width=2mm] ($(xR)-1.5*({cos(\aTheta + \aPhi)}, {sin(\aTheta + \aPhi)})$) -- ($(xR)+1.5*({cos(\aTheta + \aPhi)}, {sin(\aTheta + \aPhi)})$);
    
    % rod to left mirror
    \draw[-,black, ultra thick] (xLC) -- (xL);
    
    % rod to right mirror
    \draw[-,black, ultra thick] (xRC) -- (xR);
    
    % C coordinates
    \draw[->] (xC) -- ++({0.5*\lL},0);
        \node[anchor=west] (Cx) at ($(xC) + 0.5*(\lL, 0)$) {$ \hat{\bm{x}} $};

    % origin node
    \node[scale=0.75, black] (orig) at (0,0) {\pgfuseplotmark{*}};
        \node[anchor=south, inner sep=3mm, rotate=\aTheta] at (0,0) {$ \bm{x}_0 $};
    
    % left mirror node
    \node[scale=1.25] (Lm) at (xL) {\pgfuseplotmark{*}};
        \node[anchor=south west, inner sep=1.5mm, rotate=\aTheta] at (xL) {$ \bm{x}_L $};
        
    % right mirror node
    \node[scale=1.25] (Rm) at (xR) {\pgfuseplotmark{*}};
        \node[anchor=south east, inner sep=1.5mm, rotate=\aTheta] at (xR) {$ \bm{x}_R $};
    
    % centre of mass node
    \node[scale=0.75, black] (cm) at (xC) {\pgfuseplotmark{*}};
        \node[anchor=south, inner sep=3mm, rotate=\aTheta] at (xC) {$ \bm{x}_c $};
    
    % moveable mass node
    \node[scale=1.25, black] (mm) at (xM) {\pgfuseplotmark{*}};
        \node[anchor=south, inner sep=3mm, rotate=\aTheta] at (xM) {$ \bm{x}_m $};
    
    % phi left mirror node
    \node (Lphi) at ($(xL)-1.0*({-cos(\aTheta - \aPhi)}, {-sin(\aTheta - \aPhi)})$) {};
    
    % phi right mirror node
    \node (Rphi) at ($(xR)-1.0*({cos(\aTheta + \aPhi)}, {sin(\aTheta + \aPhi)})$) {};
    
    % Angle markers
    \pic [draw, -, "$\theta$", angle eccentricity=1.1, angle radius=1.35cm] {angle = Cx--xC--Rm};
    \pic [draw, -, "$\phi$", angle eccentricity=1.5, angle radius=0.5cm] {angle = orig--xR--Rphi};
    \pic [draw, -, "$\phi$", angle eccentricity=1.5, angle radius=0.5cm] {angle = Lphi--xL--orig};

    % Normals
    \draw [->] (xR) -- ++ ({sin(\aTheta+\aPhi)},{-cos(\aTheta+\aPhi)}) node[anchor=west] {$ \hat{\bm{n}}_R$};
    \draw [->] (xL) -- ++ ({sin(\aTheta-\aPhi)},{-cos(\aTheta-\aPhi)})  node[anchor=north] {$ \hat{\bm{n}}_L$};
    
    % xi labels
    \draw [|<->|] ($(orig) + 1.5*({sin(\aTheta)},{-cos(\aTheta)})$) -- ($(xM) + 1.5*({sin(\aTheta)},{-cos(\aTheta)})$)  node[midway, anchor = north,rotate=\aTheta] {$ \xi $};
    \draw[thin, dotted] (orig) --+ ($1.5*({sin(\aTheta)},{-cos(\aTheta)})$);
    \draw[thin, dotted] (xM) --+ ($1.5*({sin(\aTheta)},{-cos(\aTheta)})$);
    \draw [|<->|] ($(orig) + 0.75*({sin(\aTheta)},{-cos(\aTheta)})$) -- ($(xC) + 0.75*({sin(\aTheta)},{-cos(\aTheta)})$)  node[midway, anchor = north,rotate=\aTheta] {$ \varepsilon \xi $};
    \draw[thin, dotted] (xC) --+ ($0.75*({sin(\aTheta)},{-cos(\aTheta)})$);

   \definecolor{myBG}{rgb}{1,1,1}
    \node[rectangle,draw=white,fill=myBG,minimum size=2em,label={center:(b)}] at ({-\lL*cos(\aTheta)}, {\lL*sin(\aTheta)}) {};
\end{tikzpicture}
\vspace{1em}
\hrule
\vspace{1em}
\begin{tikzpicture}[scale=0.68]

    % right mirror
    \coordinate (xR) at ({\lL*sin(\aAlpha - \aTheta)}, {\lL*cos(\aAlpha - \aTheta)});
    
    % left mirror
    \coordinate (xL) at ({-\lL*sin(\aAlpha + \aTheta)}, {\lL*cos(\aAlpha + \aTheta)});
    
    % centre of mass
    \coordinate (xC) at ({-\lL*cos(\aAlpha)*sin(\aTheta)}, {\lL*cos(\aAlpha)*cos(\aTheta)});
    
     % right mirror
    \draw[draw=cyan!70!black, line width=2mm] ($(xR)-1.5*({sin(\aAlpha - \aTheta - \bPhi)}, {cos(\aAlpha - \aTheta - \bPhi)})$) -- ($(xR)+1.5*({sin(\aAlpha - \aTheta - \bPhi)}, {cos(\aAlpha - \aTheta - \bPhi)})$);
    
    % left mirror
    \draw[draw=cyan!70!black, line width=2mm] ($(xL)-1.5*({-sin(\aAlpha + \aTheta - \bPhi)}, {cos(\aAlpha + \aTheta - \bPhi)})$) -- ($(xL)+1.5*({-sin(\aAlpha + \aTheta - \bPhi)}, {cos(\aAlpha + \aTheta - \bPhi)})$);

    % rod to right mirror
    \draw[-,black, ultra thick] (0,0) -- (xR);
    
    % rod to left mirror
    \draw[-,black, ultra thick] (0,0) -- (xL);
    
    % line joining two mirrors
    \draw[-, black, dotted] (xL) -- (xR);
    
    % line dividing angle of two mirrors
    \draw[-, black, dotted] (0,0) -- (xC);
    
    % C coordinates
    \draw[->] (xC) -- ++({0.35*\lL},0);
        \node (cX) at ($(xC)+({0.35*\lL},0)$) {};
        \node[anchor=west] at (cX) {$ \hat{\bm{x}} $};

    % Nodes
    \node[scale=0.75] (orig) at (0,0) {\pgfuseplotmark{*}};
        \node[anchor=north, inner sep=2mm] at (0,0) {$ \bm{x}_{h} $};
    \node[scale=1.25] (Rm) at (xR) {\pgfuseplotmark{*}};
        \node[anchor=south east, inner sep=1.5mm, rotate=\aTheta] at (xR) {$ \bm{x}_{R} $};
    \node[scale=1.25] (Lm) at (xL) {\pgfuseplotmark{*}};
        \node[anchor=south west, inner sep=1.5mm, rotate=\aTheta] at (xL) {$ \bm{x}_{L} $};
    \node[scale=0.75] (cm) at (xC) {\pgfuseplotmark{*}};
        \node[anchor=south, inner sep=2mm, rotate=\aTheta] at (xC) {$ \bm{x}_{c} $};
    
    % left and right mirror angles
    \node (Rphi) at ($(xR)-1.5*({sin(\aAlpha - \aTheta - \bPhi)}, {cos(\aAlpha - \aTheta - \bPhi)})$) {};
    \node (Lphi) at ($(xL)-1.5*({-sin(\aAlpha + \aTheta - \bPhi)}, {cos(\aAlpha + \aTheta - \bPhi)})$) {};

    % Angle markers
    \pic [draw, -, "$\theta$", angle eccentricity=1.2, angle radius=0.85cm] {angle = cX--xC--xR};
    \pic [draw, -, "$\alpha$", angle eccentricity=1.5, angle radius=0.35cm] {angle = Rm--orig--xC};
    
    \pic [draw, -, "$\phi$", angle eccentricity=1.25, angle radius=0.75cm] {angle = orig--Rm--Rphi};
    \pic [draw, -, "$\phi$", angle eccentricity=1.25, angle radius=0.75cm] {angle = Lphi--Lm--orig};
    
    % Normals to mirrors
    \draw [->] (xR) -- ++ ({cos(\aAlpha-\aTheta-\bPhi)},{-sin(\aAlpha-\aTheta-\bPhi)}) node[anchor=west] {$ \hat{\bm{n}}_R$};
    \draw [->] (xL) -- ++ ({-cos(\aAlpha+\aTheta-\bPhi)},{-sin(\aAlpha+\aTheta-\bPhi)})  node[anchor=north] {$ \hat{\bm{n}}_L$};
    
    % Normals to arms
    \draw [->] ($0.5*(xR)$) -- ++ ({cos(\aAlpha-\aTheta)},{-sin(\aAlpha-\aTheta)})  node[anchor = north west] {$ \hat{\bm{n}}_{Rr}$};
    \draw [->] ($0.5*(xL)$) -- ++ ({-cos(\aAlpha+\aTheta)},{-sin(\aAlpha+\aTheta)})  node[anchor = north] {$ \hat{\bm{n}}_{Lr}$};

    \definecolor{myBG}{rgb}{1,1,1}
    \node[rectangle,draw=white,fill=myBG,minimum size=2em,label={center:(c)}] at ({-\lL*sin(\aAlpha + \aTheta)}, {\lL*cos(\aAlpha - \aTheta)}) {};
\end{tikzpicture}%

\caption{(a) Schematic of a simple, rigid 2D sail consisting of two mirrors at positions $\bm{x}_{L,R}$ connected by a massless rod of length $ 2 L $; the mirrors make an angle $\phi$ with the rod. Angle $\theta$ gives the orientation of the sail with respect to the incoming laser beam (red arrow). (b) Schematic of the Moveable Mass (MM) sail with a moveable point mass that is connected to the rest of the sail by damped springs. The moveable mass is at $ \bm{x}_m $ at a distance $ \xi $ from the centre of the sail $ \bm{x}_0 $; the centre of mass is displaced from $ \bm{x}_0 $ by $ \epsilon\xi $ to $ \bm{x}_c $, with $ \varepsilon $ defined in Eq.~\eqref{eq:epsilon_M}; and $\hat {\bf n}_{L,R}$ are the normals to the mirrors. (c) Schematic of Moving Arms (MA) sail in which the connecting rod has a damped hinge at the centre at $ \bm{x}_h $, $\alpha$ is the half-angle between the rods, with normals $\hat {\bf n}_{Lr}$ and $\hat {\bf n}_{Rr}$.}
\label{fig:schematics_all}
\end{figure}

Here we consider two sails with different internal degrees of freedom that have been incorporated in a rigid two-dimensional geometry (see Fig.~\ref{fig:schematics_all}(a)). We refer to these two proof-of-concept designs as Moveable Mass (MM) sail (see Fig.~\ref{fig:schematics_all}(b)) and Moving Arms (MA) sail (Fig.~\ref{fig:schematics_all}(c)). For both of these sail designs the non-relativistic equations of motion can be derived in closed form. The MM sail design is effective in damping lateral oscillations, whereas the MA sail design is not. This shows that the internal degrees of freedom need to be chosen with care. The dissipated energy adds to the energy that is inevitably absorbed by the sail from the laser beam, and which needs to be lost as thermal radiation. We show that the heat generated through damping of lateral oscillations of the sail is likely to be smaller than the absorbed laser power. We therefore conclude that the damping of undesired sail motion by dissipative internal degrees of freedom is realistic and deserves further investigation.

The outline of this paper is as follows. In Section~\ref{sec:sail_designs} we introduce the rigid sail, as well as the MM and MA sails. The equations of motion for the MM sail are obtained in Section~\ref{sec:equations_of_motion} and its linear stability is analysed in Section~\ref{sec:stability_analysis}. In Section~\ref{sec:dissipated_energy} we calculate the heat generated by the damping mechanism and compare it to the energy absorbed by the sail from the accelerating laser beam. Our numerical results are presented in Section~\ref{sec:numerical_results} and we discuss and summarise our results in Section~\ref{sec:summary}. The stability conditions for the rigid sail are given in Appendix~\ref{sec:app:characteristic_polynomial_simple}. The MA sail is discussed in Appendix~\ref{sec:app:moving_arms_sail} where we present its equations of motion and demonstrate why it does not lead to damping of the lateral oscillations of the sail. Appendix~\ref{sec:app:nominal_values} gives the nominal (dimensional) parameter values we use and Appendix~\ref{sec:app:characteristic_polynomial_mm} gives the characteristic polynomial of the MM sail.

\section{\label{sec:sail_designs} Sail designs}

To analyse the possibility of  damping through internal degrees of freedom, we start from arguably the simplest possible sail design exhibiting restoring forces towards the centre of the laser beam. Our emphasis here is to show a damped internal degree of freedom can damp any lateral perturbations and return the sail to its equilibrium point instead of focusing on practically viable mechanisms for producing restoring forces. A spherical  sail in a doughnut beam has been shown to be stable within the beam \cite{Manchester_Loeb_2017_2017ApJ...837L..20M}. Considering finite two dimensional sections of such a sphere  provides a simpler geometry, still exhibiting stability, shown in Fig.~\ref{fig:schematics_all}(a): two mirrors, each with mass $ m_1 $ located at $ {\bm x}_L $ (left mirror) and $ {\bm x}_R $ (right mirror) are connected by a massless rigid rod of length $ 2L $. The mirrors are at angle $ \phi $ with the connecting rod, which in turn has an angle $ \theta $ (orientation angle of the sail) with the $ x $-axis of the coordinate frame in which the laser is at rest -- the \textit{laser frame}. We consider perfect specular reflection by the mirrors subject to a laser beam profile, pointing along the positive $ y $-axis, given by
\begin{equation}\label{eq:beam_profile}
    I(\bm{x})
        = I_0 \left[\delta_0 + \left(\bm{x} \cdot \hat{\bm{x}} / L\right)^2\right],
\end{equation}
where $ \delta_0 \geq 0 $ is a dimensionless constant, $ \bm{x} $ is the position vector and $ \hat{\bm{x}} $ is unit vector along the $ x $-axis of the laser frame. This beam profile may be viewed as the lowest order approximation (valid for small displacements of the sail from the centre of the beam) to an annular beam \cite{Duocastella_Arnold_2012_2012LPRv....6..607D} or to a Laguerre-Gaussian mode laser beam with a nonzero orbital angular momentum number when $ \delta_0 = 0 $ \cite{Padgett_etal_2004_2004PhT....57e..35P}.

We assume that the laser is always focused on the sail (in that the lateral beam profile does not depend on the distance from the laser) and ignore relativistic effects, in particular the Doppler shift of the laser light \cite{Kulkarni_etal_2018_2018AJ....155..155K, Kipping_2017_2017AJ....153..277K, Bible_etal_2013_2013SPIE.8876E..05B,Lingam_Loeb_2020_2020ApJ...894...36L,Fuzfa_etal_2020_PhysRevResearch.2.043186} so that the momentum transfer for each photon is independent of the speed of the sail. We assume point-mirrors in the sense that the intensity of the laser is approximated to be uniform across the projected area of each mirror. The model can easily be made more realistic by replacing the point mirrors with spatially extended gratings \cite{Ilic_Atwater_2019_2019NaPho..13..289I,Srivastava_etal_2019_2019OptL...44.3082S}, but our focus here is the dissipation mechanism of the lateral motions of the sail.

The stability of this simple sail can be understood as follows: If the sail shifts to the right compared to the centre of the beam, the intensity at the right mirror increases, while that at the left mirror decreases, creating a net restoring force. Similarly, if $ \theta $ is positive, the cross section of the left mirror intercepting the beam is larger than that of the right mirror, leading to a net restoring torque. More complete linear stability analysis, see Appendix~\ref{sec:app:characteristic_polynomial_simple}, shows that this geometry is indeed stable, but not damped, i.e. it is marginally stable, similar to the more complicated designs considered in the literature, e.g. by Ilic and Atwater \cite{Ilic_Atwater_2019_2019NaPho..13..289I}.

\subsection{\label{sec:sail_designs_mm_sail} Moveable mass sail}

Figure~\ref{fig:schematics_all}(b) shows the first modification of the sail shown in Fig.~\ref{fig:schematics_all}(a) in which we introduce a damped internal degree of freedom by incorporating a moveable mass. The connecting rod now includes a housing that is located symmetrically about the centre of the sail at $ \bm{x}_0 $ between the mirrors. The housing contains a moveable point mass $ m_2 $ located at $ \bm{x}_m $ with its movement constrained along a straight line between $\bm{x}_L$ and $\bm{x}_R$. The mass is connected to the two ends of the housing by damped massless springs. The extension and compression of each spring is denoted by $ \xi $ and is measured from its equilibrium state at $ \bm{x}_0 $. The centre of mass of the sail is at $ \bm{x}_c $ at a distance $ \varepsilon \xi $ from $ \bm{x}_0 $ where
\begin{equation}\label{eq:epsilon_M}
    \varepsilon
        = m_2 / M,\quad
    \text{with}\quad
    M
        = 2 m_1 + m_2.
\end{equation}

The springs provide a restoring force on the moveable mass at $ \bm{x}_m $ towards its equilibrium position at $ \bm{x}_0 $. In the frame in which the mirrors are at rest, when the mass is perturbed it oscillates about $ \bm{x}_0 $ along the length of the connecting rod. In the laser frame the mirrors and the mass counter oscillate about the centre of mass at $ \bm{x}_c $ such that the centre of mass does not accelerate in the absence of a net nonzero external force.

As discussed in Section~\ref{sec:introduction}, the restoring force and the damping could be provided by a number of different physical mechanisms. However, our analysis is agnostic to these, provided the nature of the restoring force is not drastically different, for simplicity we refer to the restoring and damping forces as provided by springs in both MM and MA sails.

\subsection{\label{sec:sail_designs_ma_sail} Moving arms sail}

Figure~\ref{fig:schematics_all}(c) shows the second sail design that we consider. In this MA sail design, the rigid rod has a hinge halfway between the mirrors at $ \bm{x}_h $ with opening half angle $ \alpha $. The total mass for this sail is $ M = 2 m_1 $. A massless damped torsion spring at the hinge provides a torque on the arms with equilibrium at $ \alpha = \alpha_0 $. The damping in the MA sail may be provided by a magnetic damper or similar mechanisms. When the MA sail is perturbed the arms oscillate about the equilibrium opening half angle $ \alpha_0 $ and cause the centre of mass of the sail to oscillate along the line joining $ \bm{x}_c $ and $ \bm{x}_h $ in the laser frame. 

\section{\label{sec:equations_of_motion} Equations of motion}

We now derive and discuss the equations of motion of the MM sail, whereas the equations of motion for the MA sail are discussed in Appendix~\ref{sec:app:moving_arms_sail}. Our equations are written in the frame of the laser with the beam pointing along the positive $ y $-axis, as shown in Fig.~\ref{fig:schematics_all}(a).

\subsection{\label{sec:forces} Forces}

We denote the coordinates of the centre of mass of the sail as $ \bm{x}_c = [x, y] $ so that
\begin{equation}\label{eq:positions_mm}
    \bm{x}_{L,R}
        = \bm{x}_c \mp (L \pm \varepsilon \xi) \hat{\bm{r}},\quad
    \bm{x}_m
        = \bm{x}_c + (1 - \varepsilon) \xi \hat{\bm{r}},
\end{equation}
where $ \hat{\bm{r}} = [\cos\theta, \sin\theta] $ is a unit vector. The force exerted by the laser on the sail is then given by
\begin{equation}\label{eq:laser_force}
    \bm{F}_{\rm laser}
        = - \frac{2 A}{c} \sum_{i = L, R} I(\bm{x}_i) (\hat{\bm{n}}_i \cdot \hat{\bm{y}})^2 \hat{\bm{n}}_i,
\end{equation}
where $ A $ is the area of the point mirrors, $ I(\bm{x}) $ is the laser intensity, as given by Eq.~\eqref{eq:beam_profile}, and $ \hat{\bm{n}}_i $ are the unit normals to each mirror, as shown in Fig.~\ref{fig:schematics_all}(b), given by
\begin{equation}\label{eq:normals_mirrors}
    \hat{\bm{n}}_{L,R}
        = [\sin(\theta \mp \phi), - \cos(\theta \mp \phi)].
\end{equation}
In Eq.~\eqref{eq:laser_force} one power of $ (\hat{\bm{n}}_i \cdot \hat{\bm{y}}) $ arises from the projection of the area of the mirrors and the other from transfer of the momentum of photons to the mirrors. The choice of unit normals to the mirrors require that $ \phi $ and $ \theta $ must satisfy
\begin{equation}\label{eq:angle_restriction_mm}
    0 < \phi \pm \theta < \pi/2,
\end{equation}
so that laser is incident on the reflecting surface of the mirrors and the force generated is pointing towards the centre of the beam - a restoring force. This restriction is implied by Eqs~\eqref{eq:laser_force} and \eqref{eq:normals_mirrors}.

The force provided by the springs (including the damping force) on the mirrors and the moveable mass are
\begin{equation}\label{eq:spring_force_mm}
    \bm{F}_{sL,R}
        = (k \xi + \Gamma \dot{\xi}) \hat{\bm{r}}, \quad
    \bm{F}_{sm}
        = - 2 (k \xi + \Gamma \dot{\xi}) \hat{\bm{r}},
\end{equation}
where $ k $ is the spring constant of each of the springs, $ \Gamma $ is the damping coefficient and time derivative is denoted by a dot. Note that the internal forces add to zero in the inertial laser frame.

\subsection{Equations of motion}

The equations of motion of the sails may be obtained using the Newton-Euler approach or using D'Alembert's principle. We derived the equations of motion using both methods and present the latter. The more common Lagrangian or Hamiltonian approaches do not apply here as the force due to the laser does not have an associated potential. D'Alembert's principle gives the equations of motion as \cite{Goldstein_etal_2002_2002clme.book.....G}
\begin{equation}\label{eq:d_alembert}
    \frac{d}{d t} \left(\frac{\partial T}{\partial \dot{q}_j}\right) - \frac{\partial T}{\partial q_j}
        = Q_j,\quad
    Q_j
        = \sum_{i} \bm{F}_i \cdot \frac{\partial \bm{x}_i}{\partial q_j},
\end{equation}
where $ \bm{F}_i $ is the net non-constraint force on the mass with position vector $ \bm{x}_i $, $ q_j $ are generalised coordinates, $ Q_j $ are generalised forces, with the products $ q_j Q_j $ having units of energy, and $ T $ is the kinetic energy given by
\begin{equation}\label{eq:kinetic_energy}
    T
        = \frac{1}{2} \sum_{i} m_i \abs{\dot{\bm{x}}_i}^2.
\end{equation}
For the MM sail we have $ \bm{q} = (x, y, \theta, \xi) $. Using Eqs~\eqref{eq:positions_mm}--\eqref{eq:normals_mirrors} and Eqs~\eqref{eq:spring_force_mm}--\eqref{eq:kinetic_energy} we obtain the non-dimensional equations of motion of the MM sail as
\begin{widetext}
\begin{subequations}\label{eq:equations_of_motion_mm}
\begin{align}
    \label{eq:eom_mm_x}
    \ddot{\bar{x}}
        & = - \bar{I}(\bm{x}_L) \cos^2(\theta - \phi) \sin(\theta - \phi) - \bar{I}(\bm{x}_R) \cos^2(\theta + \phi) \sin(\theta + \phi),\\
    \label{eq:eom_mm_y}
    \ddot{\bar{y}}
        & = \bar{I}(\bm{x}_L) \cos^3(\theta - \phi) + \bar{I}(\bm{x}_R) \cos^3(\theta + \phi),\\
    \label{eq:eom_mm_theta}
    \ddot{\bar{\theta}}
        & = -\frac{2 \varepsilon \bar{\xi} \dot{\bar{\xi}} \dot{\bar{\theta}}}{1 + \varepsilon \bar{\xi}^2} - \frac{\cos\phi \left[(1 + \varepsilon \bar{\xi}) \bar{I}(\bm{x}_L) \cos^2(\theta - \phi) - (1 - \varepsilon \bar{\xi}) \bar{I}(\bm{x}_R) \cos^2(\theta + \phi)\right]}{(1 - \varepsilon) (1 + \varepsilon \bar{\xi}^2)},\\
    \label{eq:eom_mm_xi}
    \ddot{\bar{\xi}}
        & = \bar{\xi} \dot{\bar{\theta}}^2 - \frac{2 \bar{k} \bar{\xi} + 2 \bar{\Gamma} \dot{\bar{\xi}}}{\varepsilon (1 - \varepsilon)} - \frac{\sin\phi \left[\bar{I}(\bm{x}_L) \cos^2(\theta - \phi) - \bar{I}(\bm{x}_R) \cos^2(\theta + \phi)\right]}{1 - \varepsilon},
\end{align}
\end{subequations}
\end{widetext}
where $ I(\bm{x}_i) = I_0 \bar{I} (\bm{x}_i) $ and a bar denotes a non-dimensionalised quantity. We non-dimensionlise using mass, length and time scaling factors
\begin{equation}\label{eq:MLT_scale_factors}
    m_{\rm s}
        = M,\quad
    x_{\rm s}
        = L,\quad
    t_{\rm s}
        = \sqrt{M L c / 2 I_0 A},
\end{equation}
respectively. The scaling factors of the spring constant and damping coefficient of the MM sail are, respectively, $ k_s = m_{\rm s}/t_{\rm s}^2 $ and $ \Gamma_s = m_{\rm s} / t_{\rm s} $. Our results are independent of the specific values of the parameters that appear in Eq.~\eqref{eq:MLT_scale_factors}, however, we provide nominal values in Appendix~\ref{sec:app:nominal_values}. Henceforth, our dimensional results are obtained in terms of these nominal values.

Equations~\eqref{eq:equations_of_motion_mm} show that the $ x $, $ y $ and $ \theta $ coordinates of the MM sail do not depend directly on the restoring and damping forces provided by the spring. The damping is indirect through $ \xi $ and/or its time derivative. All terms multiplying $ \bar{I}(\bm{x}_{L,R}) $ result from the force exerted by the laser beam on the mirrors. The first term in Eq.~\eqref{eq:eom_mm_theta} can be interpreted as the Coriolis effect and arises due to time-dependence of the moment of inertia of the sail, whereas the first term in Eq.~\eqref{eq:eom_mm_xi} can be interpreted as the centrifugal effect. In the Lagrangian framework such effects result from the time dependence of the non-diagonal mass matrix in the expression of the kinetic energy of the sail. Both Coriolis and centrifugal terms arise as we are no longer dealing with a rigid body system. 

The equations of motion of the MM sail are singular when $ \varepsilon = 0 $ (for $ m_2/m_1 \to 0 $) and $ \varepsilon = 1 $ (for $ m_2/m_1 \to \infty $), but these limits are not of interest. When $ \phi \to 0 $ (mirrors are parallel to the connecting rod), Eq.~\eqref{eq:equations_of_motion_mm} simplifies considerably. In particular, the laser force is no longer present in Eq.~\eqref{eq:eom_mm_xi}. We show in Section~\ref{sec:stability_analysis} that for the sail to be stable we require $ \phi $ to be larger than some minimum value. This implies that a minimum laser force must affect $ \xi $. We also note that the equations of motion of the simple sail (see Appendix~\ref{sec:app:characteristic_polynomial_simple}), shown in Fig.~\ref{fig:schematics_all}(a), may be obtained from those of the MM sail by ignoring Eq.~\eqref{eq:eom_mm_xi} and taking the limit $ \varepsilon \to 0 $ in Eq.~\eqref{eq:eom_mm_x}--\eqref{eq:eom_mm_theta}.

Henceforth, except where explicitly stated, we work with non-dimensionalised quantities and drop the bar for simplicity in notation.

\section{\label{sec:stability_analysis} Stability Analysis}

We analyse the stability of the MM sail by linearising Eq.~\eqref{eq:equations_of_motion_mm} by expanding about the equilibrium point $ (x, \theta, \xi, \dot{x}, \dot{\theta}, \dot{\xi}) = \bm{0} $. For the sail to be (linearly) stable with damped lateral motion, we require that the eigenvalues of the coefficient matrix have negative real parts so that the general solution
\begin{equation*}
    \bm{q}
        = \sum_{s} c_s e^{\Re \lambda_s t} e^{i\Im \lambda_s t} \bm{q}_s,
\end{equation*}
decays with time. Here coefficients $ c_s $ are determined by the initial conditions, $ \bm{q}_s $ are eigenvectors corresponding to eigenvalues $ \lambda_s $ and $ s $ runs over all non-trivial eigenmodes.

Linearising the equations of motion of the MM sail results in the characteristic polynomial given in Appendix~\ref{sec:app:characteristic_polynomial_mm} [Eq.~\eqref{eq:app:characteristic_mm}]. There are six non-trivial eigenvalues and two trivial ones. The trivial eigenvalues result because the equations of motion do not depend on coordinate $ y $ and its time derivative. This is a consequence of our assumption that the laser beam profile has no dependence on $ y $ and our neglect of the Doppler shift of the laser wavelength (so that there is no dependence on $ \dot{y} $). The characteristic polynomial of the MM sail cannot be solved analytically. We therefore employ the Routh-Hurwitz criterion \cite{Gradshteyn_Ryzhik_2007_2007tisp.book.....G} to obtain a set of necessary and sufficient conditions for all solutions of Eq.~\eqref{eq:app:characteristic_mm} to have nonzero negative real parts. These conditions are
\begin{equation}\label{eq:stability_conditions_mm}
    \Gamma > 0, \quad 
    k > k_{\rm min}\quad
    {\rm and}\quad
    \phi_{\rm min} < \phi < \pi/2, 
\end{equation} 
where
\begin{equation}\label{eq:kphi_min}
\begin{split}
    k_{\rm min} 
        & = (1 + \delta_0) \varepsilon^2 \cos^2\phi \sin\phi,\\
    \phi_{\rm min} 
        & = \cos^{-1}\left[1 - \frac{9 + 6 \delta_0 + \delta_0^2}{19 - 4 \varepsilon + 14 \delta_0 + 3 \delta_0^2}\right]^{1/2},
\end{split}
\end{equation}
with $ 0 < \varepsilon < 1 $ and $ \delta_0 \geq 0 $. We observe, as expected,  that a nonzero damping coefficient $ \Gamma $ is necessary for the sail to be stable (and damped). The minimum value of $ \phi $ increases with increasing $ \varepsilon $ and decreases with increasing $ \delta_0 $. This results, respectively, in increasing and decreasing the restoring component ($ x $-component) of the laser force while doing the opposite for the net force. The minimum spring constant $ k $ increases with minimum laser intensity $ \delta_0 $ and mass ratio $ \varepsilon $ and decreases with increasing $ \phi $ (increasing laser restoring force and decreasing total laser force). The conditions for minimum $ k $ and $ \phi $ ensure that a sufficient restoring force is provided on the moveable mass and on the sail, respectively.

Equation~\eqref{eq:stability_conditions_mm} gives a wide range of parameters over which all eigenmodes are damped. That is, for which any lateral displacement of the sail (including rotations) are damped provided the perturbations from equilibrium are sufficiently small for nonlinear effects to be insignificant.

\begin{figure}[hbt]
    \centering
    \includegraphics[width=\columnwidth]{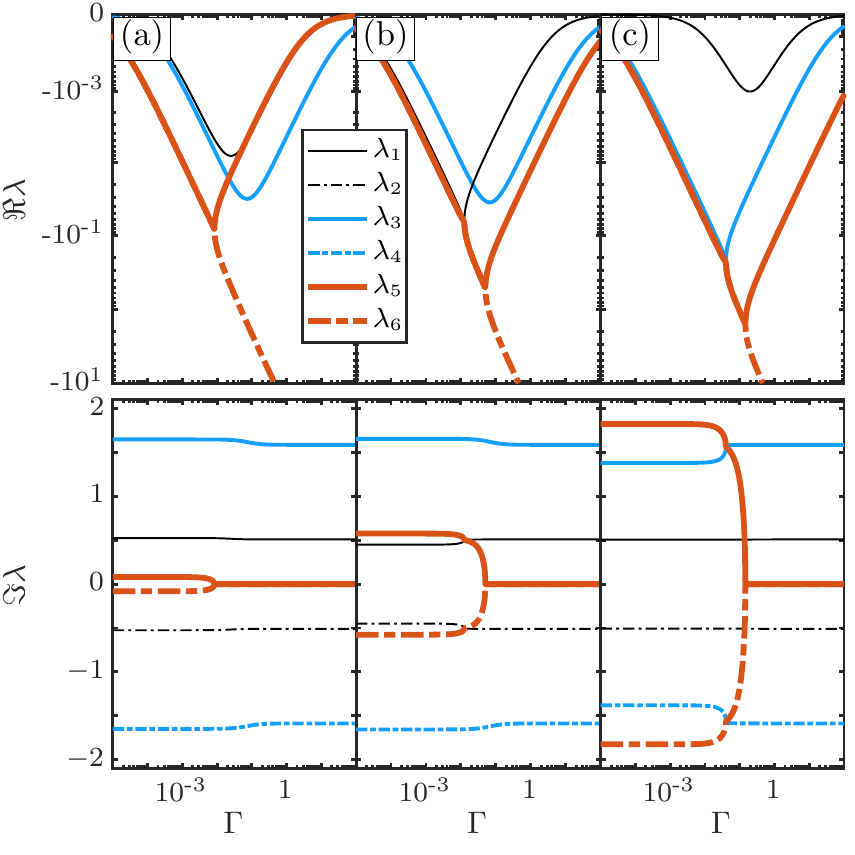}
    \caption{Eigenvalues of MM sail. Real (top row) and imaginary (bottom row) parts versus $ \Gamma $ for $ \delta_0 = 0.1 $, $ \varepsilon = 0.1 $, $ \phi/\phi_{\rm min} = 1.15 $ and $ k/k_{\rm min} = 1.1 $ [column (a)], $ 4.6869 $ [column (b)] and $ 33.717 $ [column (c)]. Different eigenvalue pairs are shown using different colours, line thicknesses and styles (see legend). }
    \label{fig:eigenvalues}
\end{figure}

Figure~\ref{fig:eigenvalues} shows the real (top row) and imaginary (bottom row) parts of the non-trivial eigenvalues as a function of damping coefficient $ \Gamma $ for parameter values given in the figure caption. The vertical axis of the top row uses a bi-symmetric logarithmic scaling with range $ \abs{\Re \lambda} \leq 1 $ represented over eight decades. The eigenvalues are either purely real or occur in complex conjugate pairs, as expected. As $ k/k_{\rm min} $ is increased, the $ \Im \lambda_{1,2,3,4} $ are relatively unaffected, with $ 0.5 \lesssim \abs{\Im \lambda_{1,2}} \lesssim 0.52 $ and $ 1.6 \lesssim \abs{\Im \lambda_{3,4}} \lesssim 1.7 $, while $ \abs{\Im \lambda_{5,6}} $ increase with $ k/k_{\rm min} $ from zero at $ k/k_{\rm min} = 1 $. However, $ \Im \lambda_{5,6} $ is nonzero only for small $ \Gamma $ up to the bifurcation point of $ \lambda_{5,6} $ where $ \lambda_{5,6} $ becomes purely real. For $ k/k_{\rm min} \gtrsim 4.7 $, the bifurcation occurs at $ \Gamma \approx \sqrt{k / k_{\rm min}}/40 $ for the parameters used in the figure; the scaling $ \propto \sqrt{k / k_{\rm min}} $ is maintained when $ \delta_0 $, $ \varepsilon $ and $ \phi $ are changed. Past the bifurcation point, the magnitude of $ \Re \lambda_5 $ decreases towards zero while that of $ \Re \lambda_6 $ continues to increase with $ \Gamma $. The real parts of $ \lambda_{1,2} $ and $ \lambda_{3,4} $ remain relatively unaffected as $ k/k_{\rm min} $ is increased except near values where they coalesce in the complex plane with $ \lambda_{5,6} $ when $ k/k_{\rm min} \approx 4.69 $ and $ \approx 33.7 $, respectively, as seen in the columns (b) and (c) of Fig.~\ref{fig:eigenvalues}.

The magnitude of the troughs of the real parts of the relevant eigenvalues increase to a maximum at coalescence points and then decrease and move to larger $ \Gamma $ with increasing $ k / k_{\rm min} $. These coalescence points indicate resonance points between these eigenmodes where they share a common dissipation rate and oscillation frequency. This implies that for damping lateral motions of the sail the optimal values of $ k / k_{\rm min} $ and $ \Gamma $ are possibly those that correspond to the intersections of $ \lambda_{1,2} $ and $ \lambda_{5,6} $ as eigenmodes corresponding to $ \lambda_{1,2} $ have the slowest decay rates. This choice also avoids (potential) high frequency oscillations of the sail corresponding to nonzero $ \Im \lambda_{5,6} $ when $ k/k_{\rm min} $ is large; specially when past the intersections of $ \lambda_{3,4} $ and $ \lambda_{5,6} $. However, the precise choice of optimal values of the parameters leading to fastest decay rate is a nontrivial exercise as the coordinates of coalescence points in the $ (k, \Gamma) $ plane  also depend on $ \delta_0 $, $ \varepsilon $ and $ \phi $. In general, the coalescence points move to lower $ k $ and higher $ \Gamma $ coordinates with increasing $ \delta_0 $ and $ \phi $ and to lower $ k $ and higher $ \Gamma $ with $ \varepsilon $.

The bifurcation of $ \lambda_{5,6} $ as a function of $ k $ and $ \Gamma $ are similar to the appearance of the exceptional points in PT-symmetric systems in quantum mechanics and optics \cite{Bender_Boettcher_1998_1998PhRvL..80.5243B,Miri_Alu_2019_Mirieaar7709,Suchkov_etal_2016_2016LPRv...10..177S}, for example that of a passive, lossy plasmonic-dielectric waveguide model \cite{Tuniz_etal_2019_PhysRevLett.123.213903}. The spring constant $ k $ plays the role of the coupling constant, with $ k_{\rm min} $ being its critical value, while $ \Gamma $ corresponds to the loss in the supporting waveguide. Previous reports in passive optical two-mode systems~\cite{Tuniz_etal_2019_PhysRevLett.123.213903} showed that the exceptional point is associated with the largest overall dissipation. In this mechanical six-mode system, the topology of the eigenspace is significantly more complicated and the choice of optimal parameters is not obvious.

When the conditions given in Eq.~\eqref{eq:stability_conditions_mm} are satisfied, we are guaranteed to have an asymptotically stable equilibrium. The neighbourhood of the equilibrium where linear stability is applicable could be very small as (typically small factors) $ k $ and $ \Gamma $ multiply linear terms in Eq.~\eqref{eq:equations_of_motion_mm}, i.e. there is the potential danger of exiting the linear regime for seemingly small perturbations about the equilibrium. We must choose $ \delta_0 $, $ \varepsilon $, $ \phi $, $ k $ and $ \Gamma $ judiciously if a specific damping rate of the lateral oscillations of the sail is desired and/or if lateral perturbations in a specific neighbourhood of the equilibrium are to be damped; however, the latter is not always guaranteed.

\begin{figure*}
    \centering
    \includegraphics[width=\textwidth]{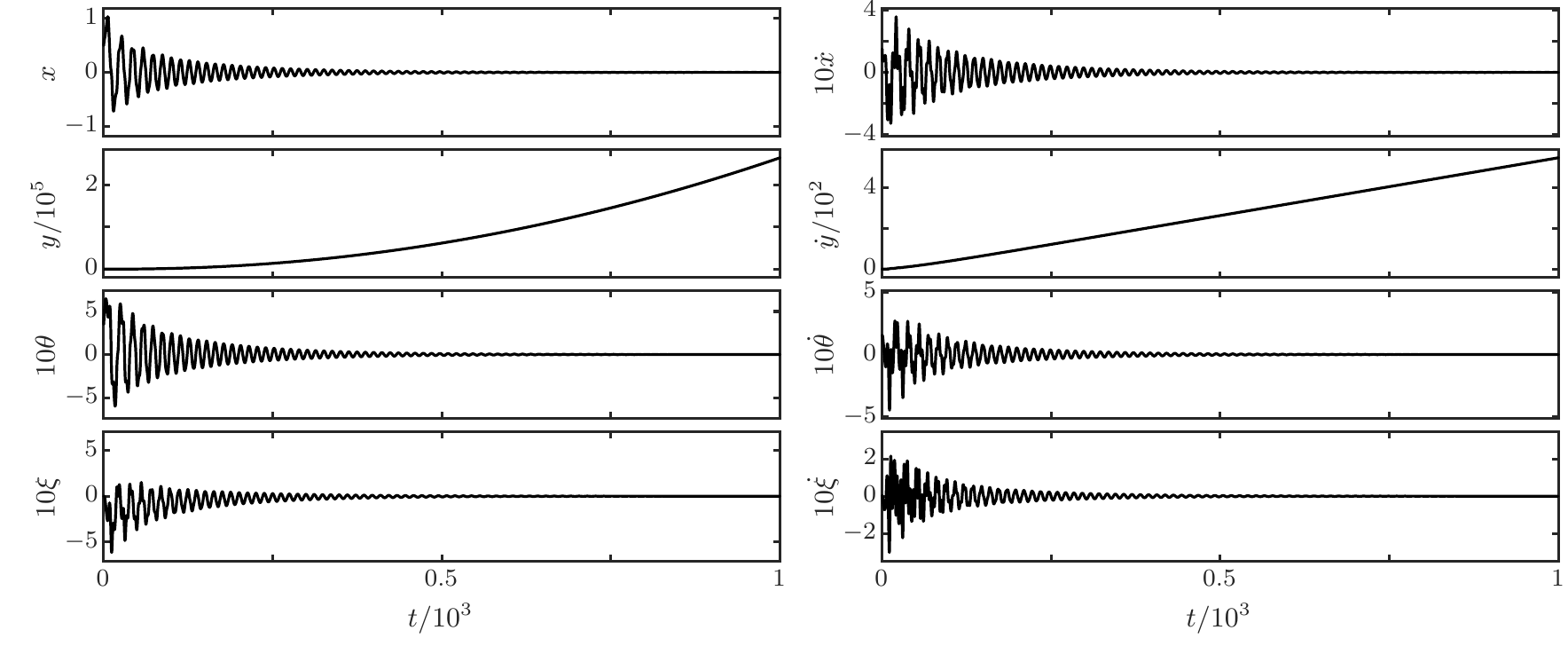}
    \caption{Elements of $ \bm{q} $ and $ \dot{\bm{q}} $ as a function of time $ t $ for parameter values given in the column (a) of Fig.~\ref{fig:eigenvalues} with $ \Gamma = 0.04 $ and for initial conditions $ (x_0, \theta_0, \dot{x}_0, \dot{\theta}_0) = (0.5, 0.35, 0.15, 0.15) $.}
    \label{fig:coord_plots}
\end{figure*}

Figure~\ref{fig:coord_plots} shows $ \bm{q} = (x, y, \theta, \xi) $ and $ \dot{\bm{q}} $ versus time $ t $ for the same parameter values used in column (a) of Fig.~\ref{fig:eigenvalues} with $ \Gamma = 0.04 $ and initial conditions $ (x_0, \theta_0, \dot{x}_0, \dot{\theta}_0) = (0.5, 0.35, 0.15, 0.15) $, calculated by integrating the full non-linear equations of motion Eq.~\eqref{eq:equations_of_motion_mm}. The vertical and horizontal axes have been scaled for convenience as shown. The choice of $ (x_0, \theta_0) = (0.5, 0.35) $ is arbitrary and the additional perturbations in $ \dot{x}_0 $ and $ \dot{\theta}_0 $ increase $ x $ and $ \theta $ in the absence of damping. For comparison Ilic and Atwater~\cite{Ilic_Atwater_2019_2019NaPho..13..289I} use $ (x_0, \theta_0, \dot{x}_0, \dot{\theta}_0) = (0.5, 0.1, 0, 0) $ to obtain undamped oscillations of the sail they consider. It is evident that all lateral motions of the sail are damped over time. The motion along the $ y $-axis has (almost) constant acceleration and is decoupled in this (mostly) linear regime with $ y(t) \approx \left[(1 + \delta_0) \cos^3\phi\right] t^2 $. This decoupling implies that the damping mechanism only affects the lateral components of the sail motion, as desired.

The lateral oscillations of the sail are irregular in the initial phase of the motion and become more regular with increasing time as the fastest decaying modes damp out. The quantities $ x $, $ \theta $, $ \dot{x} $ and $ \dot{\theta} $ reach their maximum (absolute) values at times $ t \approx 6.1 $, $ 3.2 $, $ 21 $ and $ 11 $, respectively, and reduce to $ 1/e $ of their maximum values after \textit{the damping time} $ \Delta t_{\rm 1/e} \approx 58 $, $ 98 $, $ 91 $ and $ 74 $, respectively. For the nominal parameter values in Appendix~\ref{sec:app:nominal_values}, these correspond to dimensional times $ \Delta t_{\rm 1/e} \approx 0.17 $\,s, $ 0.29 $\,s, $ 0.27 $\,s and $ 0.22 $\,s, respectively. We obtain these estimates by fitting a curve to the envelope of the absolute values of $ \bm{q} $ and $ \dot{\bm{q}} $. The decay time for the lateral perturbations of the sail thus quite short compared to the nominal acceleration time of $ \approx 900 $\,s. 

The longest surviving oscillations have (non-dimensional) angular frequency $ \omega \approx 0.52 $ corresponding to $ \Im \lambda_{1,2} $ in column (a) of Fig.~\ref{fig:eigenvalues}. This corresponds to a dimensional angular frequency of approximately $ 68\,{\rm rad\, s}^{-1} $. The other four eigenmodes have dimensional oscillation frequencies of $ \omega_{3,4} \approx 560\,{\rm rad\, s}^{-1} $ and $ \omega_{5,6} \approx 28\,{\rm rad\, s}^{-1} $. The oscillations of the sail due to $ \omega_{5, 6} $ may be avoided through appropriate choice of parameters, as discussed above, and $ \omega_{3,4} $ may be tuned to a desired value by changing the sail parameters that affect scaling factor of time, $ t_{\rm s} $.

In contrast, the stability analysis of the MA sail (Appendix~\ref{sec:app:moving_arms_sail}) shows that, unlike the MM sail, at best it can have two damped eigenmodes and four undamped ones. The undamped modes of the MA sail do not dissipate when excited and hence the MA sail is not asymptotically stable. Furthermore, any deviations from a narrow range of parameter values lead to emergence of growing modes and instability.

\section{\label{sec:dissipated_energy} Dissipated energy}

We demonstrated that all lateral perturbations in coordinates $ x $, $ \theta $ and their time derivatives for the MM sail are damped. The dissipated energy is additional heat that can only be lost through thermal radiation. Cooling of the sail is a difficult task in itself \cite{Ilic_etal_2018_2018NanoL..18.5583I} and as such the damping mechanism must produce as little additional heat as possible. Here we discuss the rate of heat production by the damping mechanism of the MM sail and compare it to the expected heating of the sail through absorption of the laser light.

The total energy incident on the sail from the laser beam in time $ t $ is
\begin{equation}\label{eq:energy_absorbed}
    E_L(t)
        = - \frac{c t_{\rm s}}{2L} \sum_{i = L, R} \int_0^{t} I(\bm{x}_i) \left(\hat{\bm{n}}_i \cdot \hat{\bm{y}}\right) dt'.
\end{equation}
An actual sail absorbs energy $ a_\alpha E_L(t) $, where $ a_\alpha $ is the total absorption coefficient. We wish to compare the heat generated by damping of the lateral oscillations of the sail to this absorbed energy from the laser. The heat generated by damping is given by the work done by the moveable mass against the damping force
\begin{equation}\label{eq:heat_generated}
    W_\Gamma(t)
        = 2 \Gamma \int_0^t \dot{\xi}^2 dt,
\end{equation}
where we use Eq.~\eqref{eq:spring_force_mm}. The ratio of thermal energy due to damping of the lateral motion to absorbed laser energy is $ r_h = r_H / r_\alpha $ where
\begin{equation}
    r_H
        = \frac{W_\Gamma(t)}{E_L(t)}.
\end{equation}

\begin{figure}[hbt]
    \centering
    \includegraphics[width=\columnwidth]{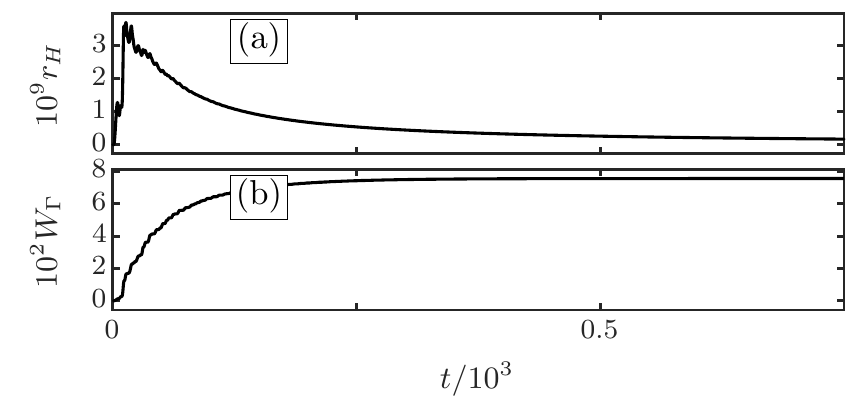}
    \caption{(a) $ r_H $ and (b) $ W_\Gamma $ versus time $ t $ for the same parameter values and initial conditions as in Fig.~\ref{fig:coord_plots}.}
    \label{fig:heat_plots}
\end{figure}

Figure~\ref{fig:heat_plots} shows (a) $ r_H $ and (b) $ W_\Gamma $ versus time for the parameters in Fig.~\ref{fig:coord_plots}. The total heat produced due to damping is approximately $ 7.55 \times 10^{-2} $ which is smaller than the total laser energy landing on the sail by a factor of $ 7.9 \times 10^{9} $ as shown by the plot of $ r_H $. The rate of heat generation through damping is largest in the early stages of motion where lateral perturbations are largest. As the perturbations damp, the rate of heat generated drops as well. At its peak, the heat generated through damping is equivalent to a total absorption factor $ \approx 3.71 \times 10^{-9} $. Reasonable total absorption for candidate sail materials is $ a_\alpha \gtrsim 10^{-8} $ \cite{Atwater_etal_2018_2018NatMa..17..861A} which is significantly higher than the heat generated through damping of the lateral oscillations of the sail. In addition to this, the set of initial conditions chosen in Figs~\ref{fig:coord_plots} and \ref{fig:heat_plots} are quite extreme. In reality, we expect milder perturbations of the sail and hence a reduced rate of heat generation due to the damping of the lateral perturbations of the sail.

\section{\label{sec:numerical_results} Numerical results}

The results presented in the previous section are for a single set of values of parameters $ \delta_0 $, $ \varepsilon $, $ \phi $, $ k $ and $ \Gamma $ and for a single set of initial conditions $ (x_0, \theta_0, \dot{x}_0, \dot{\theta}_0) $. Here we analyse the rate of heat generation and total heat generated, as well as the damping time of the perturbations, as the parameters are varied. We choose a default set of parameters $ \delta_0 = 0 $, $ \varepsilon = 0.1 $, $ \phi / \phi_{\rm min} = 1.05 $, $ k / k_{\rm min} = 1.1 $ and $ \Gamma = 10^{-2} $ with, using Eq.~\eqref{eq:kphi_min}, $ \phi_{\rm min} \approx 0.769 $ and $ k_{\rm min} \approx 3.45 \times 10^{-3} $ for the chosen values of $ \delta_0 $ and $ \varepsilon $, and vary one parameter at a time. We consider the range of parameter values
\begin{equation}
\begin{gathered}
    0 \leq \delta_0 \leq 0.3,\quad
    0.1 \leq \varepsilon \leq 0.4,\quad
    0.01 \leq \Gamma \leq 0.1,\\
    1.1 \leq k/k_{\rm min} \leq 5.6,\quad
    1.05 \leq \phi/\phi_{\rm min} \leq 1.2,
\end{gathered}
\end{equation}
and for each parameter we take 7 equally spaced samples except for $ \Gamma $ which is sampled logarithmically. We divide the initial condition ranges
\begin{equation}
    \abs{x_0} \leq 0.15, \quad
    \abs{\theta_0} \leq 0.15,\quad
    \abs{\dot{x}_0} \leq 0.15,\quad
    \abs{\dot{\theta}_0} \leq 0.15,
\end{equation}
into 7 equal parts for a total of $ 2401 $ sets of initial conditions. We solve the equations of motion by using the default values for all parameters except one which is varied over the range of values described above. For each set of parameter values, we solve the equations of motion for all initial conditions and estimate the damping time (see Section~\ref{sec:stability_analysis}). We also calculate the maximum of $ r_H $ and the total heat $ W_\Gamma $ generated through damping.

\begin{figure}[hbt]
    \centering
    \includegraphics[width=\columnwidth]{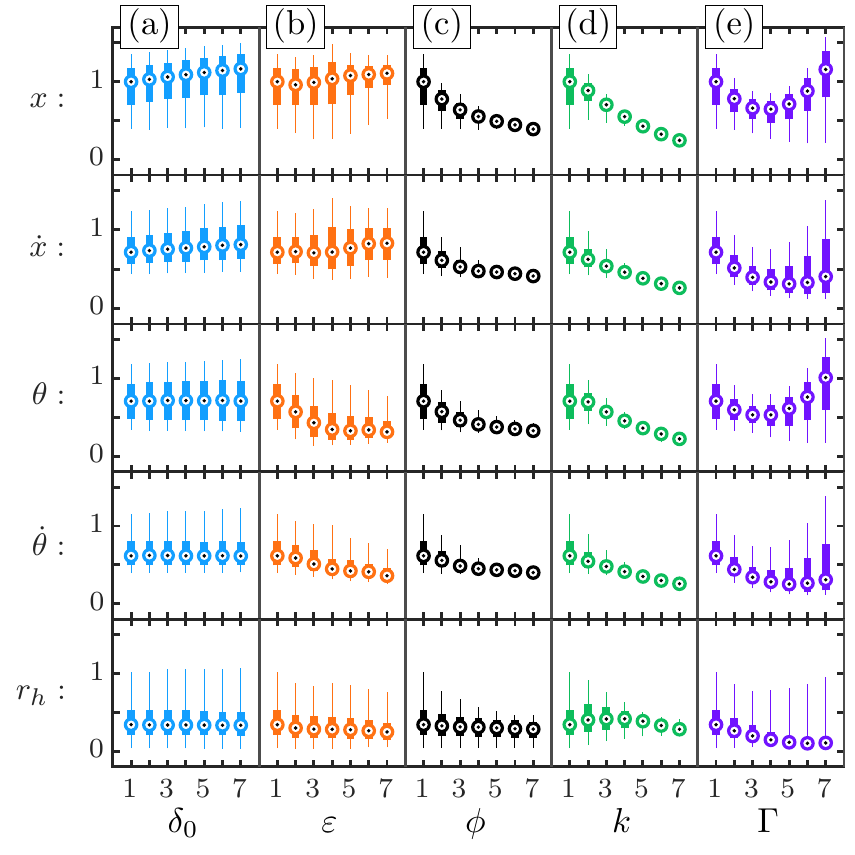}
    \caption{The MM sail is stable for a wide range of parameters and initial perturbations: Box-and-whiskers plots of the damping time of $ x $ (first row), $ \dot{x} $ (second row), $ \theta $ (third row), $ \dot{\theta} $ (fourth row) and $ r_H $ (fifth row) while varying the parameters (horizontal axes) $ \delta_0 $ [column (a)], $ \varepsilon $ [column (b)], $ \phi $ [column (c)], $ k $ [column (d)] and $ \Gamma $ [column (e)]. Bulls-eye: median; thick portion of the box: $25^{\rm th}-75^{\rm th}$ percentile range; thin lines: $5^{\rm th}-95^{\rm th}$ percentile range.}
    \label{fig:time_stat_results}
\end{figure}

Figure~\ref{fig:time_stat_results} shows the time taken (vertical axes) for each of  $ x $ (first row), $ \dot{x} $ (second row), $ \theta $ (third row), $ \dot{\theta} $ (fourth row) and $ r_H $ (fifth row) to reduce to $ 1/e $ of their peak values as we vary the parameters (horizontal axes) $ \delta_0 $ [column (a)], $ \varepsilon $ [column (b)], $ \phi $ [column (c)], $ k $ [column (d)] and $ \Gamma $ [column (e)] while keeping the rest fixed. Each data point is a box-and-whisker plot of the estimated time for the $ 2401 $ set of initial conditions with the bulls-eye indicating the median, the thick portion of the box indicating the $25^{\rm th}$ to $75^{\rm th}$ percentile ranges and the thin lines (the whiskers) indicating the $ 5^{\rm th}-95^{\rm th}$ percentile ranges. In each cell the left-most data point is for the default set of parameter values. All time values have been scaled by the median of the damping time of $ x $ for the default parameter values. The numbers $ 1 $ to $ 7 $ on the horizontal axes indicate the index of the value of the corresponding parameter below them. All axes are linear except for the horizontal axes of the column (e) which is logarithmic owing to the logarithmic sampling of $ \Gamma $.

Varying $ \delta_0 $ has little effect on the damping times. This is not unexpected as $ \delta_0 $ varies the intensity, and hence the force of the laser uniformly. This is counter-acted by varying the minimum spring constant $ k_{\rm min} $ and minimum mirror angle $ \phi_{\rm min} $ accordingly to adjust the restoring forces on the mirrors and the moveable mass, Eq.~\eqref{eq:stability_conditions_mm}. Increasing $ \varepsilon $ causes $ \theta $ and $ \dot{\theta} $ to damp more rapidly while producing the opposite effect for $ x $ and $ \dot{x} $. Increasing $ \phi $ and $ k $ both result in reduced damping time for all quantities. Changing $ \phi $ affects the restoring force provided by the laser on the mirrors while varying $ k $ affects the restoring force on the moveable mass. The interplay of these two restoring forces gives the coupling between the damped degree of freedom and other coordinates and their velocities. Increasing $ \Gamma $ initially results in a reduction of the damping time due to increased damping force, however, as $ \Gamma $ is increased further the damping times increase. This is as expected and evident from Fig.~\ref{fig:eigenvalues} which shows that the magnitude of the real part of all eigenvalues, except for $ \lambda_6 $, increase with $ \Gamma $ and then decrease. Finally, the time taken for the maximum heat produced to reduce to $ 1/e $ of its maximum value is affected only meaningfully by $ k $ and $ \Gamma $.

The results presented in Fig.~\ref{fig:time_stat_results} show only slices of the full parameter space and its effect on damping behaviour. Choosing the optimal set of parameter values for a specific metric is nontrivial, as discussed above, and a full analysis of the damping times over the full parameter space is beyond the scope of this paper. However, the inclusion of the damped internal degree of freedom undoubtedly damps the lateral oscillations of the sail in a reasonably small time compared with the total acceleration time of the sail.

\begin{figure}[hbt]
    \centering
    \includegraphics[width=\columnwidth]{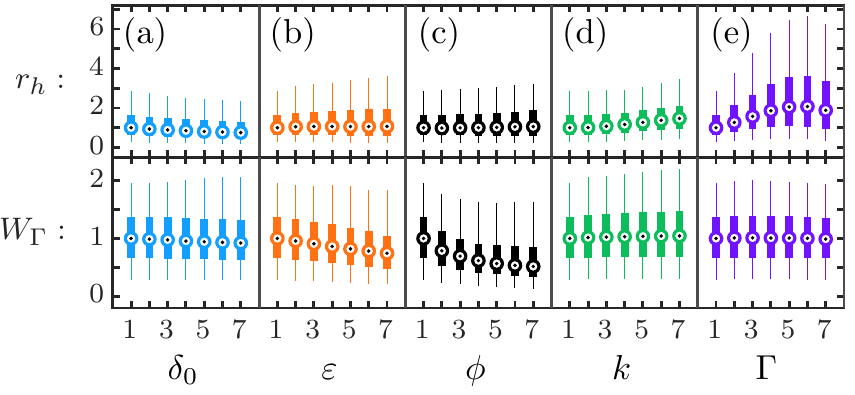}
    \caption{Box-and-whiskers plots of values of $ r_H $ (top row), scaled by $ 5.25 \times 10^{-11} $, and $ W_\Gamma $ (bottom row), scaled by $ 3.10 \times 10^{-2} $, as we vary the parameters (horizontal axes) $ \delta_0 $ [column (a)], $ \varepsilon $ [column (b)], $ \phi $ [column (c)], $ k $ [column (d)] and $ \Gamma $ [column (e)]. The legend is the same as in Fig.~\ref{fig:time_stat_results}.}
    \label{fig:heat_stat_results}
\end{figure}

Figure~\ref{fig:heat_stat_results} shows box-and-whiskers plots of the values of $ r_H $ (top row) and $ W_\Gamma $ (bottom row) as described above. We observe that rate of heat production depends most strongly on $ k $ and $ \Gamma $ and the total amount of heat produced varies appreciably only when $ \phi $ is changed. Larger $ k $ and $ \Gamma $ results in greater rate of heat production, unsurprisingly, as $ \Gamma $ is directly responsible for heat production and larger $ k $ means larger $ \dot{\xi} $ and hence a larger rate of heat production.

\section{\label{sec:summary} Discussion and conclusions}

The prospect of interstellar travel is an idea that excites and motivates the pursuit of new discoveries in numerous areas of science and technology. Our investigations are certainly still in their infancy as we move towards a viable solution. In this paper we address one of the central challenges: stability of the sail during its acceleration phase. We modify a simple sail to include a damped internal degree of freedom and demonstrate that the lateral oscillations of the sail may be damped effectively leading to sail stability. We give the criteria for stability and demonstrate that all small lateral perturbations may be damped effectively for a wide range of sail parameter values. We consider the extra heat generated from the damping of the lateral motions and show that it is likely to be much smaller than the heat generated through absorption of the laser light by the sail.

We propose and analyse two sails with internal damped degrees of freedom and analyse their ability to damp lateral oscillations of the sail. The designs we discussed are merely two examples of a much larger class. However they are opposites in that in the MM design the centre of mass moves in the direction that connects the two mirrors, whereas in the MA design the centre of mass moves perpendicular to this direction. Therefore, even though the MM and MA designs are but two implementations of damped internal degrees of freedom, they each are arguably representative of a much larger set of designs. We may tentatively conclude that, perhaps not surprisingly, the damping of lateral motion requires an internal degree of freedom that can move in the same direction to linear order. This is so even though all of the sails' degrees of freedom are coupled in the equations of motion (see Eqs~\eqref{eq:equations_of_motion_mm} and~\eqref{eq:app:equations_of_motion_ma}). 

In choosing optimal parameters one might be tempted to consider the equivalent of critical damping, which corresponds to the exceptional point in PT-symmetric systems \cite{Fernandez_2018_2018EJPh...39d5005F}. However, the condition for critical damping pertains to a single pair of modes ($\lambda_{5,6}$ in Fig.~\ref{fig:eigenvalues}), whereas the other modes are not critically damped. Since the residual motion is determined by the most weakly damped modes ($\lambda_{1,2}$ in Fig.~\ref{fig:eigenvalues}), it is may be more advantageous to consider the coalescence points, where accidental degeneracies can lead to the fastest damping. 

While our analysis is two-dimensional, it can in principle be straightforwardly generalised to three dimensions. This allows not only sideways translation and rotational motion, but also rotation about the axis of the laser beam. The main issue is how to generalise the geometry. One could consider an additional damped mass-spring system orthogonal to the present one, or perhaps $ n $ of these at mutual angles of $ \pi/n $. All of the geometries shown in Fig.~\ref{fig:damping_schemes} are examples of such generalisations. However, this is outside the scope of the present work.

\begin{acknowledgments}

We thank Christopher Poulton, Simon Fleming, Peter Tuthill, Sergio Leon-Saval, Stefano Palomba and Iver Cairns for continued helpful discussions and suggestions. The work was supported by the Science Foundation for Physics and is a Grand Challenge project at the School of Physics, University of Sydney.

\end{acknowledgments}

\appendix

\section{\label{sec:app:characteristic_polynomial_simple} Simple sail}

The characteristic polynomial of the simple sail may be obtained from linearising its equations of motion or from Eq.~\eqref{eq:app:characteristic_mm} by taking the limits $ \bar{k} / \varepsilon \to 0 $, $ \bar{\Gamma} / \varepsilon \to 0 $ and $ \varepsilon \to 0 $ as
\begin{equation}\label{eq:app:characteristic_simple}
    P(\lambda)
        = \lambda^2 \left[\lambda^4 + \Lambda_4 \lambda^2 + \Lambda_2\right],
\end{equation}
with
\begin{equation}
    \Lambda_4
        = 4 (2 + \delta_0) \cos^2\phi \sin\phi,\quad
    \Lambda_2
        = 8 (1 + \delta_0) \cos^6\phi,
\end{equation}
and with nontrivial solutions
\begin{equation}
    \lambda 
        = \pm \left\{-\frac{\Lambda_4}{2} \left[1 \pm \Delta^{1/2}\right]\right\}^{1/2},\quad
    \Delta
        = 1 - \frac{4 \Lambda_2}{\Lambda_4^2}.
\end{equation}
The system is marginally stable (i.e. all eigenvalues are purely imaginary) when $ 0 \leq \Delta \leq 1 $, which is satisfied provided we have
\begin{equation}
    \phi_{\rm min} \leq \phi \leq \frac{\pi}{2},\quad
    \phi_{\rm min}
        = \tan^{-1} \left[\frac{2 (1 + \delta_0)}{(2 + \delta_0)^2}\right],
\end{equation}
and the system is unstable otherwise. 

\section{\label{sec:app:moving_arms_sail} Moving arms sail}

Here we derive the equations of motion of the MA sail, shown in Fig.~\ref{fig:schematics_all}(c) and discussed in Section~\ref{sec:sail_designs_ma_sail}, and analyse its stability.

\subsection{\label{sec:app:equations_of_motion} Equations of motion}

The equations of motion of the the MA sail may be obtained in a similar way to that of the MM sail (see Section~\ref{sec:equations_of_motion}). Here we have
\begin{equation}\label{eq:app:positions_ma}
    \bm{x}_{L,R}
        = \bm{x}_c \mp L \sin\alpha \hat{\bm{r}},
\end{equation}
and force due to the laser is given by Eq.~\eqref{eq:laser_force} but with
\begin{equation}\label{eq:app:normals_mirrors}
    \hat{\bm{n}}_{L,R}
        = [\mp \cos(\alpha \pm \theta - \phi), - \sin(\alpha \pm \theta - \phi)],
\end{equation}
as shown in Fig.~\ref{fig:schematics_all}(c). The force exerted by the torsion spring on the two mirrors of MA sail is given by
\begin{equation}\label{eq:app:spring_force_ma}
    \bm{F}_{sL,R}
        = [k (\alpha_0 - \alpha) - \Gamma \dot{\alpha}] \hat{\bm{n}}_{Lr,Rr},
\end{equation}
where the unit normals to the arms, as shown in Fig.~\ref{fig:schematics_all}(c), are given by
\begin{equation}\label{eq:app:normals_arms}
    \hat{\bm{n}}_{Lr,Rr}
        = [\mp \cos(\alpha \pm \theta), - \sin(\alpha \pm \theta)].
\end{equation}

The non-dimensional equations of motion for MA sail are then obtained as
\begin{subequations}\label{eq:app:equations_of_motion_ma}
\begin{align}
    \label{eq:app:eom_ma_x}
    \ddot{\bar{x}}
        & = \bar{I}(\bm{x}_L) \sin^2(\alpha + \theta - \phi) \cos(\alpha + \theta - \phi)\nonumber\\
        &\quad - \bar{I}(\bm{x}_R) \sin^2(\alpha - \theta - \phi) \cos(\alpha - \theta - \phi)\nonumber\\
        &\qquad + \sin\alpha \sin\theta \left[\bar{k} \left(\alpha_0 - \alpha\right) - \bar{\Gamma} \dot{\bar{\alpha}}\right],\\
    \label{eq:app:eom_ma_y}
    \ddot{\bar{y}}
        & = \bar{I}(\bm{x}_L) \sin^3(\alpha + \theta - \phi) + \bar{I}(\bm{x}_R) \sin^3(\alpha - \theta - \phi)\nonumber\\
        &\quad - \sin\alpha \cos\theta \left[\bar{k} \left(\alpha_0 - \alpha\right) - \bar{\Gamma} \dot{\bar{\alpha}}\right],\\
    \label{eq:app:eom_ma_theta}
    \ddot{\bar{\theta}}
        & = - 2 \cot\alpha\, \dot{\bar{\alpha}}\, \dot{\bar{\theta}} - \frac{\sin(\alpha - \phi)}{\sin\alpha}\left[\bar{I}(\bm{x}_L) \sin^2(\alpha + \theta - \phi)\right.\nonumber\\
        &\quad\left. - \bar{I}(\bm{x}_R) \sin^2(\alpha - \theta - \phi)\right],\\
    \label{eq:app:eom_ma_alpha}
    \ddot{\bar{\alpha}}
        & = \tan\alpha\, (\dot{\bar{\alpha}}^2 + \dot{\bar{\theta}}^2) + \left[\bar{k}(\alpha_0 - \alpha) - \bar{\Gamma} \dot{\bar{\alpha}}\right]\nonumber\\
        &\quad - \frac{\cos(\alpha - \phi)}{\cos\alpha}\left[\bar{I}(\bm{x}_L) \sin^2(\alpha + \theta - \phi)\right.\nonumber\\
        &\qquad\left. + \bar{I}(\bm{x}_R) \sin^2(\alpha - \theta - \phi)\right],
\end{align}
\end{subequations}
where we non-dimensionalise using
\begin{equation}\label{eq:MLT_scale_factors_ma}
    m_{\rm s}
        = m_1,\quad
    x_{\rm s}
        = L,\quad
    t_{\rm s}
        = \sqrt{M L c / I_0 A},
\end{equation}
and note that $ k_s = m_{\rm s} x_{\rm s} / t_{\rm s}^2 $ and $ \Gamma_s = m_{\rm s} x_{\rm s} / t_{\rm s} $ for the MA sail. Again, we drop the bar on the non-dimensionalised parameters for simplicity in notation.

Unlike Eq.~\eqref{eq:equations_of_motion_mm} we observe that for the MA sail the spring force couples directly to the $ x $- and $ y $-coordinates in addition to the internal degree of freedom $ \alpha $. The Coriolis effect (first term of Eq.~\eqref{eq:app:eom_ma_theta}) and centrifugal effect (first term of Eq.~\eqref{eq:app:eom_ma_alpha}) are again present. The singularity in the equations of motion of the MA sail at $ \alpha = \pi / 2 $ can be removed if we add a nonzero mass at the hinge, as discussed above. This additional mass would regularise the equations of motion without affecting the stability of the system. We choose $ 0 < \alpha_0 < \pi $ but sufficiently away from $ \pi/2 $ so that the oscillations of the arms about $ \alpha = \alpha_0 $ do not cross $ \alpha = \pi/2 $.

\subsection{\label{sec:app:stability_analysis} Stability Analysis}

Linearising the equations of motion about the equilibrium point $ (x, \theta, \alpha, \dot{x}, \dot{\theta}, \dot{\alpha}) = (0, 0, \alpha_0, 0, 0, 0) $ gives a characteristic polynomial of the form
\begin{equation}\label{eq:app:characteristic_ma}
    P(\lambda)
        = \lambda^2\left[\lambda^2 + \Lambda_{11} \lambda + \Lambda_{10}\right] \left[\lambda^4 + \Lambda_{21} \lambda^2 + \Lambda_{22}\right],
\end{equation}
which may be solved directly. The only set of solutions where there are no eigenvalues with a nonzero positive real part are
\begin{subequations}\label{eq:app:eigenvalues_ma}
\begin{align}
    \lambda_{1,2}
        & = - \frac{1}{2} \Gamma \left\{1 \pm \sqrt{1 - 4 (k + \Delta_1) / \Gamma^2}\right\},\\
    \lambda_{3,4,5,6}
        & = \pm i \Delta_2 \left\{1 \pm \sqrt{\Delta_3}\right\}^{1/2},
\end{align}
\end{subequations}
subject to
\begin{equation}\label{eq:app:stability_conditions_ma}
    k + \Delta_1 \geq 0,\quad
    \Delta_2 \geq 0,\quad
    \Delta_3 \geq 0,
\end{equation}
where
\begin{subequations}\label{eq:app:stab_conds_ma_Deltas}
\begin{align}
    \Delta_1
        & = \left(\frac{3}{2} + 2 \delta_0\right) \sin(2\alpha_0 - 3\phi) - \sin(4\alpha_0 - 3\phi)\nonumber\\
        & \quad + \frac{1}{2} \sin(2\alpha_0 - \phi) + 2 (1 + \delta_0) \frac{\cos^2\phi}{\cos^2 \alpha_0} \sin\phi,\\
    \Delta_2
        & = 4 \left(1 + \frac{\delta_0}{2 \sin^2\alpha_0}\right)\sin\alpha_0 \sin^2(\alpha_0 - \phi) \cos(\alpha_0 - \phi),\\
    \Delta_3
        & = 1 - \frac{\tan^2(\alpha_0 - \phi) \left[1 + \delta_0 \csc^2\alpha_0\right]}{2 (1 + \delta_0 \csc^2\alpha_0 / 2)^2}.
\end{align}
\end{subequations}
Two of the eigenvalues have negative real parts and four eigenvalues are purely imaginary. We note that the damping coefficient $ \Gamma $ only enters the expression for $ \lambda_{1,2} $ and does not appear in the expressions for $ \lambda_{3,4,5,6} $. Mathematically, the characteristic polynomial factors indicate that there is block-diagonalisation, and hence the dissipation does not couple to the all degrees of freedom. Thus not all perturbations of the MA sail damp over time. The equations of motion of the MA sail are more nonlinear compared to those of the MM sail as is evident from Eq.~\eqref{eq:equations_of_motion_mm} and \eqref{eq:app:equations_of_motion_ma}. This is likely to lead to faster growth of perturbations of the sail due to nonlinear effects. Therefore, designs based on MA sail are not viable if one intends to damp out all lateral perturbations of the sail.

\section{\label{sec:app:nominal_values} Nominal parameter values}

To obtain a nominal set of parameter values we consider a 3D sail with mass $ m_{\rm sail} = 10^{-3} $\,kg and surface area $ A_{\rm sail} = 10\,{\rm m}^2 $, subject to a laser beam with uniform intensity profile, that has an intensity such that it accelerates the sail, under perfect specular reflection at normal incidence and ignoring relativistic effects, to $ 0.2 c $ in $ t_{\rm accel} = 15 $ minutes. The sail then has surface mass density $ \rho_{\rm sail} = m_{\rm sail} / A_{\rm sail} = 10^{-4} $\,kg/m. Assuming similar acceleration for the MM sail with $ A = 10^{-2}\,{\rm m}^2 $, $ L = 1 $\,m, $ m_1 = A \rho_{\rm sail} $, $ \delta_0 = 0.1 $, $ \varepsilon = 0.1 $, $ \phi = 1.15 \phi_{\rm min} $ and $ \ddot{y} = 2 (1 + \delta_0) \cos^3\phi $ requires that $ I_0 \approx 3.9 \times 10^9\,{\rm W/m}^2 $. These give
\begin{equation}
\begin{gathered}
    m_{\rm s}
        \approx 2.22 \times 10^{-6}\,{\rm kg},\quad
    x_{\rm s}
        = 1\,{\rm m},\quad
    t_{\rm s}
        \approx 2.92 \times 10^{-3}\,{\rm s},\\
    k_{\rm s} 
        \approx 0.260\, {\rm kg/s}^2,\quad
    \Gamma_{\rm s} 
        \approx 7.6 \times 10^{-4}\,{\rm kg/s}.
\end{gathered}
\end{equation}

\section{\label{sec:app:characteristic_polynomial_mm} MM sail - characteristic polynomial}

The characteristic polynomial of the MM sail is
\begin{equation}\label{eq:app:characteristic_mm}
    P(\lambda)
        = \lambda^2 \left[\lambda^6 + \Lambda_5 \lambda^5 + \Lambda_4 \lambda^4 + \Lambda_3 \lambda^3 + \Lambda_2 \lambda^2 + \Lambda_1 \lambda + \Lambda_0\right],
\end{equation}
where
\begin{subequations}\label{eq:app:eom_coefficiences_mm}
\begin{align}
    \Lambda_0
        & = \frac{16 (1 + \delta_0) \cos^6\phi \left\{k - (1 + \delta_0) \varepsilon^2 \cos^2\phi \sin\phi\right\}}{\varepsilon (1 - \varepsilon)^2},\\
    \Lambda_1
        & = \frac{16 \Gamma (1 + \delta_0) \cos^6\phi}{\varepsilon (1 - \varepsilon)^2},\\
    \Lambda_2
        & = \frac{8 (1 + \delta_0) \cos^4\phi \left\{(1 + \delta_0 \varepsilon) \cos^2\phi - (1 + \delta_0) \varepsilon\right\}}{(1 - \varepsilon)^2}\nonumber\\
        &\qquad + \frac{8 k (2 + \delta_0 - \varepsilon) \cos^2\phi \sin\phi}{\varepsilon (1 - \varepsilon)^2},\\
    \Lambda_3
        & = \frac{8 \Gamma (2 + \delta_0 - \varepsilon) \cos^2\phi \sin\phi}{\varepsilon (1 - \varepsilon)^2},\\
    \Lambda_4
        & = \frac{2 k + 4(2 + \delta_0) \varepsilon \cos^2\phi \sin\phi}{\varepsilon (1 - \varepsilon)},\\
    \Lambda_5
        & = \frac{2 \Gamma}{1 - \varepsilon}.
\end{align}
\end{subequations}

% \nocite{*}

\bibliography{lightsail}% Produces the bibliography via BibTeX.
	
\end{document}